\pdfoutput=1
\documentclass[%
 reprint,
superscriptaddress,
 amsmath,amssymb,
 aps,
floatfix,
]{revtex4-2}

\usepackage{float}
\usepackage{graphicx}% Include figure files
\usepackage{dcolumn}% Align table columns on decimal point
\usepackage{bm}% bold math
\usepackage[usenames,dvipsnames]{color} % useful color pack

\begin{document}

\title{Empirical analysis of congestion spreading in Seoul traffic network}% Force line breaks with \\

\author{Jung-Hoon Jung}
\affiliation{%
 Department of Physics, University of Seoul, Seoul 02504, Republic of Korea,
}%
\author{Young-Ho Eom}%
 \email{yheom@uos.ac.kr}
\affiliation{%
 Department of Physics, University of Seoul, Seoul 02504, Republic of Korea,
}%
\affiliation{Natural Science Research Institute, University of Seoul, Seoul 02504, Republic of Korea}
\affiliation{Urban Big data and AI Institute, University of Seoul, Seoul 02504, Republic of Korea}

\date{\today}% It is always \today, today,

\begin{abstract}
Understanding how local traffic congestion spreads in urban traffic networks is fundamental to solving congestion problems in cities. In this work, by analyzing the high resolution data of traffic velocity in Seoul, we empirically investigate the spreading patterns and cluster formation of traffic congestion in a real-world urban traffic network. To do this, we propose a congestion identification method suitable for various types of interacting traffic flows in urban traffic networks. Our method reveals that congestion spreading in Seoul may be characterized by a tree-like structure during the morning rush hour but a more persistent loop structure during the evening rush hour. Our findings suggest that diffusion and stacking processes of local congestion play a major role in the formation of urban traffic congestion.
\end{abstract}

\maketitle

%%%%%%%%%%%%%%%%%%%%% body.tex %%%%%%%%%%%%%%%%%%%%%%%%%%%%%%%%%%%%%
%% abstract
%% figure caption

\newcommand{\anote}[1]{\textcolor{red}{#1}}

\section{Introduction}
Understanding the functionality and congestion of urban traffic networks is a crucial problem as these networks serve as the blood vessels of cities~\cite{mahmassani2013Urban, wen2017Understanding,ccolak2016understanding,olmos2018macroscopic}. Since an urban traffic network is a connected network of local traffic flows on individual roads in a city, the functionality of the network relies on not only these flows but also the interactions between the flows. 

A remarkable phenomenon owing to such interactions is the spreading of local traffic congestion across the network, creating macroscopic congestion such as clusters of congested traffic flows~\cite{zhang2019scale, ambuhl2023Understanding, Nguyen2017Discovering}. A percolation-based approach was recently proposed to investigate how such congested clusters form as the number of congested traffic flows increases~\cite{li2015percolation, zeng2019switch, zeng2020multiple,kwon2023,hamedmoghadam2021percolation,cogoni2021stability}. This approach revealed that the ways that congested clusters form (or functional clusters break up) during rush hour and non-rush hour can be qualitatively different~\cite{zeng2019switch,zeng2020multiple,kwon2023}. Other studies used models of cascading failure or epidemic spreading to identify the patterns of congestion spreading in urban traffic networks~\cite{chen2022quasicontagion,Zhao2016,Saberi2020}. A recent work~\cite{Zhao2016} showed that traffic congestion, using the Motter-Lai cascading failure model~\cite{Motter2002}, in Beijing spreads radially from the center of the initial congestion with an approximately constant velocity. Another work~\cite{Saberi2020}, using the susceptible-infected-recovered model of epidemic spreading~\cite{pastor2015epidemic}, showed that the growth and decay patterns in the number of congested roads in several cities are well described by this simple epidemic model. However, to get deeper insight into the development and unfolding of urban traffic congestion, we need to ask how local congestion actually spreads and how this leads to the formation of macroscopic congestion in urban traffic networks.

To empirically address these questions, we need to resolve the following two issues about congestion identification in urban traffic networks. First, we need to determine consistently whether a given traffic flow is congested or not, regardless of the various types of roads that exist in the networks. Many existing studies use a global threshold value of flow velocity for congestion identification. However, a global threshold may not be effective when each flow has its own characteristics such as average velocity, velocity variance, or velocity distribution. For example, a single threshold value suitable for flows on highways may not be suitable for flows on other types of roads. Alternatively, the fundamental diagram~\cite{geroliminis2008Existence,geroliminis2011hysteresis,ambuhl2023Understanding} may identify the functional state of traffic flows but it requires not only flow velocity data but also vehicle density data, which are usually quite difficult to obtain. Second, we need to take into account the fact that the functionality of a traffic flow depends on not only the quality of the flow itself but also the quality of the flows on the neighboring roads, as urban traffic flows are not just an ideal gas of traffic flows but a network of traffic flows connected by the underlying road network. Considering neighboring flows is also helpful in a practical sense because most urban traffic data, collected from floating vehicles by the Global Navigation Satellite System (GNSS), are error-prone~\cite{wang2016Mining, kan2019Traffic, qin2020Building}. 

In this paper, we propose a congestion identification method suitable for various types of interacting traffic flows in urban traffic networks to resolve the above two issues. The proposed method allows us to determine the state of traffic flows by collapsing their behavior onto a single type of statistical distribution and considering the states of their neighboring flows. With the proposed method, we analyze high-resolution traffic velocity data in Seoul to empirically investigate how local congestion spreads and forms congestion clusters in the Seoul traffic network. We revealed that congestion spreading in Seoul is characterized by a tree-like structure during the morning rush hour but a more persistent loop structure during the evening rush hour, indicating that urban traffic congestion arises through the diffusion and stacking processes of local congestion.

\section{Data and Methods}\label{spa}
% Detailed scope and interest of our paper 
\begin{figure}
    \centering
    \includegraphics[width=0.92\columnwidth]{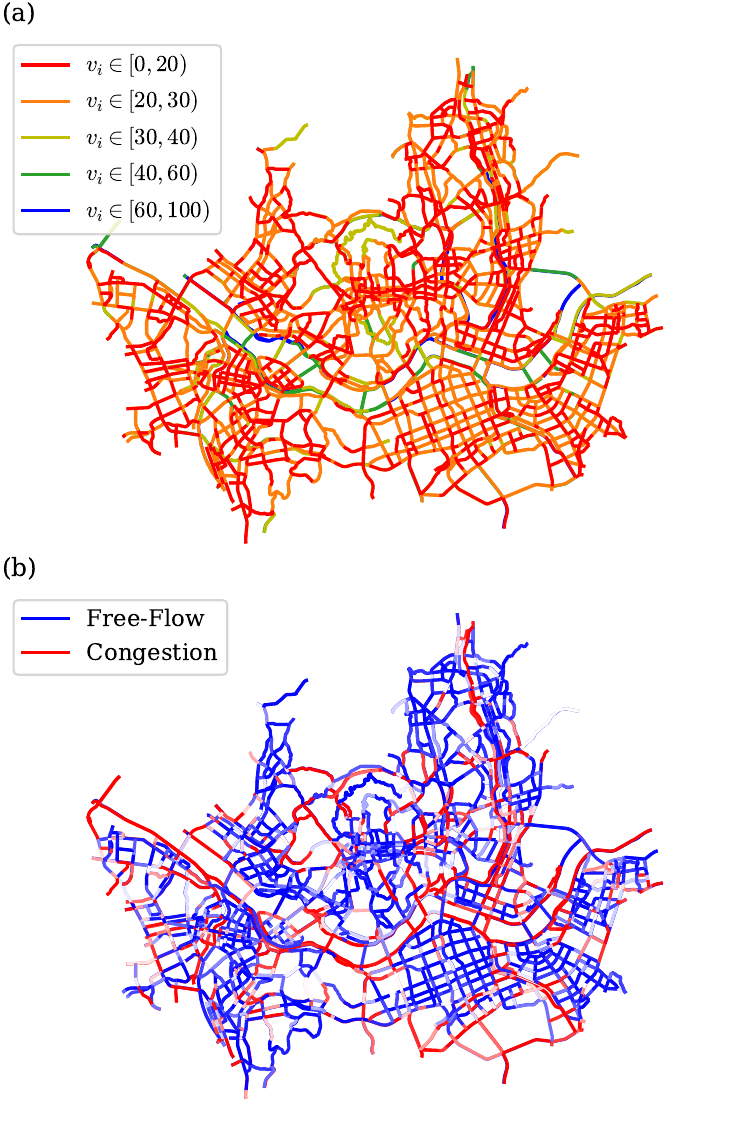}
    \caption{Spatial representation of traffic flows on the Seoul road network. 
    The time of the plot is 08:30 A.M. on Dec. 1st. 2020, which shows the pattern of the morning rush hour.
    Each road is drawn using the geometry data provided by the TOPIS and colored according to its properties.
    (a) Spatial distribution of traffic flows with their velocity. 
    Because different flows have different velocity limits and different properties, 
    it is difficult to compare the state of different traffic flows using the raw value of their velocity.
    (b) Spatial distribution of traffic flows with the resulting state vector that is calculated and calibrated by the state propagation algorithm.
    Each flow is colored blue and red if it is in a free-flow state and a congested state, respectively.
    Shade of each color is based on the converged state $s^*_i$ of each flow $i$.
    %The clearer color means the clearer state of that flow.
    }
    \label{fig:Seoul}
\end{figure}

\subsection{Data and traffic network construction}
% Dataset description & network construction 
First, we prepared the set of roads and the averaged velocities of traffic flows on these roads in Seoul, which are provided by the authorities of Seoul Transport Operation and Information Service (TOPIS)~\cite{TOPIS}. 
The traffic system of Seoul provides traffic services for more than 8 million commuters in the Seoul metropolitan area, which suffers from severe congestion.
The velocity of the traffic flow on each road was estimated from taxi GNSS data and averaged over 5-minute intervals, with a total of 288 (data points/day) $\times$ 60 (workdays) = 17,280 data points from December 2020 to February 2021.
Fig.~\ref{fig:Seoul}(a) shows an example of the velocity data at 8:30 A.M. on Dec. 1st. 2020 by assigning each flow a raw value of velocity. 
To reduce the temporal fluctuation, we adopted a 30-minute moving average velocity.

% Flow-to-flow network construction and data annealing
Next, we built a flow-to-flow network where traffic flows on individual roads correspond to nodes. A directional edge from flow $i$ to flow $j$ is created if these flows are directly connected by the underlying road network and vehicles can travel from flow $i$ to flow $j$ given the direction of travel of the flows.  
This network construction is equivalent to the traditional dual network construction, except that it additionally considers the direction of travel of the vehicles~\cite{anez1996Dual, wen2017Understanding, feng2019Identification, guo2019identifying}.
Every connection in this network represents a real-world interaction between different traffic flows, so the resulting network does not simply mimic the appearance of the underlying road network but represents the actual organization of traffic flows.
Furthermore, when we extract a subgraph of traffic congestion, such a subgraph shows the organization of congested flows in terms of connected components.
This not only makes the results easier to interpret than conventional methods but also makes it convenient to consider the influence of neighboring flows. 

% data filtering
To make the flow-to-flow network a connected system, we extracted the weakly connected components from the network without missing data points and filtered out traffic flows that were not included in the largest connected component.
This filtering is negligible and does not affect the subsequent results.
Finally, we obtained the Seoul traffic flow network covering the entire city with 4,711 flows (nodes) and 10,724 connections (edges). 

\begin{figure*}
    \centering
    \includegraphics[width=0.92\textwidth]{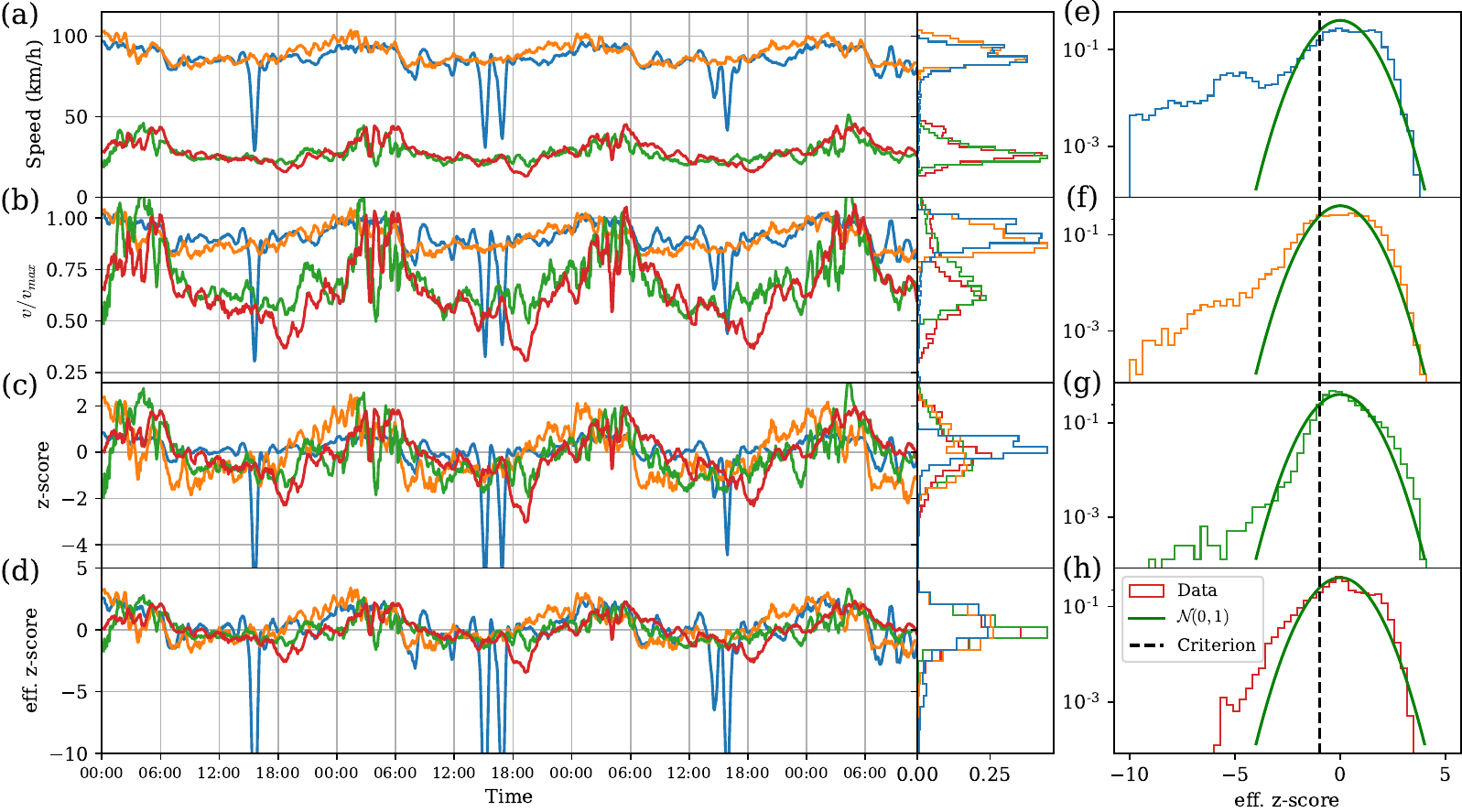}
    \caption{Samples of traffic velocities and their normalized values in various ways.
    We sampled two traffic flows on highways and two traffic flows on local roads. We colored each flow the same color in each panel.
    Each row on the left shows the pattern over time of each traffic index for the sampled traffic flow: (a) velocity, (b) relative velocity, (c) z-score, and (d) effective z-score.
    The histogram of each row on the left represents the probability density function of each traffic index for the sampled traffic flow. 
    Figures (e)-(h) on the right show the effective z-score distribution of each traffic flow and the normal distribution $\mathcal{N}(0,1)$ as a guide.
    Each row is equivalent to the same colored histogram in (d). 
    The criterion for identifying congestion is indicated by the dashed black line, which is $-1\sigma$ from an undisturbed normal distribution.
    }
    \label{fig:norm}
\end{figure*}

\subsection{Congestion as an anomalously low functional state}
% definition of Congestion as an anomalous signal
Many researchers have tried to determine whether a traffic flow is congested with various traffic indices~\cite{li2015percolation, zhang2019scale, zeng2019switch, Saberi2020, ambuhl2023Understanding}.
A typical traffic index in previous studies is the velocity ratio with the daily maximum velocity~\cite{li2015percolation, zeng2019switch, Saberi2020}, which is a simple and powerful method for normalization between different traffic flows. 
However, this method has a crucial limitation, which is that the determination of congestion only relies on the ratio of instant velocity to the maximum velocity not on the velocity distribution. 
Thus congestion identification based on this index may be biased or inconsistent as the index loses significant information about flow velocity during the day.

We suggest a more consistent way of congestion identification for each flow by leveraging its velocity distribution. We regard the congestion of a traffic flow as its failure, which is an anomalously low-velocity state. We assume that its velocity distribution obtained from data has two parts: a part that is disturbed by congestion (and therefore skewed toward low velocities) and a part that is not disturbed~\cite{dey2006Speed, jun2010Understanding, iannone2013SPEED, zefreh2020Distribution}.

% Flow normalization: effective log z score
First, we estimate the velocity distribution of each traffic flow undisturbed by its congestion in order to distinguish these two parts.
We assumed that the undisturbed velocity distribution of a traffic flow follows a lognormal distribution, considering that velocity fluctuations may affect the ratio rather than just the value.
Because the congestion of a traffic flow affects only its lower velocity, one would expect the right side of its whole velocity distribution to be undisturbed by congestion.
In practice, we effectively estimated the mean $\mu_i^{eff}$ and standard deviation $\sigma_i^{eff}$ of the undisturbed velocity distribution of a given traffic flow $i$ from its velocity sequence data $v_i(t)$ (i.e., data of its velocity time series) as below,
%equation of effective z-score normalization
\begin{equation}\label{eq:meanSD}
\begin{split}
\mu_i^{eff} &= \log m_i,\quad
    \sigma_i^{eff} = \frac{\log P95_i - \mu_i}{2},            
\end{split}
\end{equation}
where $m_i$ denotes the median (i.e., the 50 percentile) of the velocity sequence of traffic flow $i$ and $P95_i$ means the 95 percentile, regarded as the maximum velocity $v_{max}$.
Note that the effective standard deviation is approximated as half the log difference between $P95_i$ and $m_i$.
We define the effective z-score $z_i(t)$ of velocity sequences $v_i(t)$ for each traffic flow $i$ as below,
\begin{equation}\label{eq:effz}
    z_i(t) = \frac{\log v_i(t) -\mu_i^{eff}}{\sigma_i^{eff}}.            
\end{equation}
We use this normalized index to define the congestion of flow $i$ at time $t$ as its low functional state such that $z_i(t)$ is lower than a given threshold (i.e., the state with an anomalously low velocity that would be difficult to observe in the undisturbed velocity distribution).

% validation of normalization with an effective lognormal distribution
Fig.~\ref{fig:norm} shows a sample of velocities represented by flows on highways (high average velocity) and flows on local roads (low average velocity), and the results of several normalization methods of the sampled data.
As shown in Fig.~\ref{fig:norm}(a), each traffic flow has its own velocity distribution with distinguishable fluctuation and average velocity, so a direct comparison between the raw velocity data is not meaningful. 
Fig.~\ref{fig:norm}(b) represents the relative velocity $(r = v/v_{max})$. With this simplest normalization, most velocity sequences are scaled to the range of $[0,1]$ so that the sequences can be compared, but you can see that the resulting distributions still have different means and standard deviations.
Thus, if one determines whether a traffic flow is congested by comparing a single threshold value with its relative velocity, the threshold value suitable for flows on highways may not be suitable for flows on other types of roads as the bias depending on the road type (e.g., highway or local road) still remains in the relative velocity.
Another traffic index is the z-score, which is obtained by dividing the differences between a given sequence and its mean by its standard deviation (Fig.~\ref{fig:norm}(c)).
Note that in this case the mean values of all sequences are almost identical as they are close to 0, but the magnitude of the variation still depends on the road type, which affects the identification of urban congestion with a fixed threshold.
This is because the calculation of the mean and standard deviation was disturbed by congestion, suggesting that a typical z-score is not free from the effects of congestion. 
Finally, in the case of the effective z-score we proposed (Fig.~\ref{fig:norm}(d)), one can see that not only the mean but also the variance are well aligned, meaning that all the data are well described by a single type of distribution. 
These results can be seen as validating our assumptions of the velocity distribution of each traffic flow as well as the identification of congestion using a specific threshold. 
Therefore, we adopt the effective z-score to estimate the performance (i.e., quality of service) of each traffic flow and use it to identify congestion with a given threshold.

% Congestion identification on single road level
For a given congestion threshold $h$, the state $s^{(0)}_i(t)$ of traffic flow $i$ is initially estimated by the tangent hyperbolic function as below, 
\begin{equation}\label{eqn:initial}
    s^{(0)}_i(t) = \tanh{(z_i(t) +h)},
\end{equation}
where $z_i(t)$ denotes the effective z-score of the velocity of traffic flow $i$ at time $t$.
The negative and positive indicators represent a congested state and a free-flowing state, respectively.
We set the congestion threshold $h$ as 1, which means that a traffic flow that shows a performance lower than one standard deviation of the daily typical performance is considered congested (Figs.~\ref{fig:norm}(e)-(f)).
Because this identification originated from the estimated undisturbed distribution of each traffic flow, 
the resulting vector is less affected by the statistical properties of each flow and thus represents its dynamical state well.
% This kind of nonlinear activation is inspired by deep learning algorithm~\cite{scarselli2009Graph, lv2015Traffic, tian2018LSTMbased}, which preserves the information about the state of each flow into binary as well, so it is useful to calibrate the state of each traffic flow with its neighboring flows.
This kind of nonlinear activation is inspired by deep learning algorithm~\cite{scarselli2009Graph, lv2015Traffic, tian2018LSTMbased}, which preserves the information about the state of each flow into binary as well, so it is useful to limit the strength of a calibration of the state of each traffic flow with its neighboring flows.

\subsection{Congestion identification with neighboring flows: State Propagation}
In terms of traffic capacity, congested flows are in a state where they are unable to handle the loaded traffic, 
so they can make all routes that include them worse~\cite{carmona2020cracking}.
This means that the impact of a congested traffic flow is not confined to itself, but also affects the wider cluster of flows that are connected by the underlying road network. 
Therefore, when determining the state of a traffic flow, we should consider also the states of the neighboring flows. 

For example, if a given traffic flow is in a free-flowing state but the connected flows are all congested, then the flow can be considered congested. 
Conversely, even if the current performance of a traffic flow has dropped slightly, 
it should still be considered in a free-flowing state if its neighboring flows are in good condition. 
Estimating the state of nodes (i.e., flows) in this way facilitates tracking congestion spreading and brings us more robust results from the noise in the velocity data.

% State Propagation Model 
We implemented the above approach in an algorithm we call the \emph{state propagation algorithm} (SPA).
In detail, it updates the state vector ${s}_i^{(n+1)}(t)$ by calibrating the performance of node $i$ in the flow-to-flow network using all the states ${s}_j^{(n)}(t)$ of its outgoing neighbors, which is written as follows,
\begin{equation}\label{eqn:spa}
    {s}_i^{(n+1)}(t) = \tanh \left(J \sum_j \frac{{A}_{ij} {s}_j^{(n)}(t)}{k_i} + {z}_i(t) + h \right),
\end{equation}
where ${A}$ denotes the adjacency matrix of the flow-to-flow network, 
$J$ is the overall strength of the calibration by the state propagation,
$k_i(=\sum_j A_{ij})$ denotes the out-degree of node $i$,
${z}_i(t)$ is the effective z-score introduced above, 
and $h$ represents the congestion threshold. 
We set $J$ and $h$ as 1, 
which means the propagation affects a calibration of one standard deviation to the neighboring flows as maximum,
and a traffic flow that shows a performance lower than one standard deviation of the daily typical performance is considered as congested.
Note that the flow-to-flow network is a unidirectional graph, so the state propagation is also unidirectional.
We repeat this process for sufficiently large $n$ and use the converged state ${s}^*_i$ as a flow state. 
Finally, we calculated the congestion indicator $c_i(t)$ using the converged state ${s}_i^*$,
\begin{equation}
    c_i(t) = \Theta(-s^*_i(t)),
\end{equation}
where $\Theta(\cdot)$ denotes the Heaviside step function.
A stability analysis of the converged state is described in Appendix A.
If $s^*_i$ is positive, the state of flow $i$ is identified as free-flowing, if not, congested.

% properties of resulting vectors
Fig.~\ref{fig:Seoul}(b) shows the congestion identification result $s_i^*(t)$ of the SPA for the velocity data represented in Fig.~\ref{fig:Seoul}(a).
The blue and red flows indicate free-flowing and congested states, respectively, which show the clear structural patterns of urban congestion.
This is because the gradual propagation of the information of the local traffic state reinforces the structure of the underlying road network.

We believe that the result shows robust structures in the temporal evolution of urban congestion, 
specifically, even when there are so many cars on all the roads that they start to slow down, but congestion has not yet occurred.
For the local flow level identification in Eq.~(\ref{eqn:initial}), very small noise can make a big difference in the pattern because the traffic index is close to the decision boundary of congestion.
However, in the SPA, these small noises could be ignored due to the propagation effect of neighboring states.
In this sense, the SPA provides adequate results for analyzing the evolutionary pattern of congestion in urban traffic networks.

\begin{figure*}
    \centering
    \includegraphics[width=\textwidth]{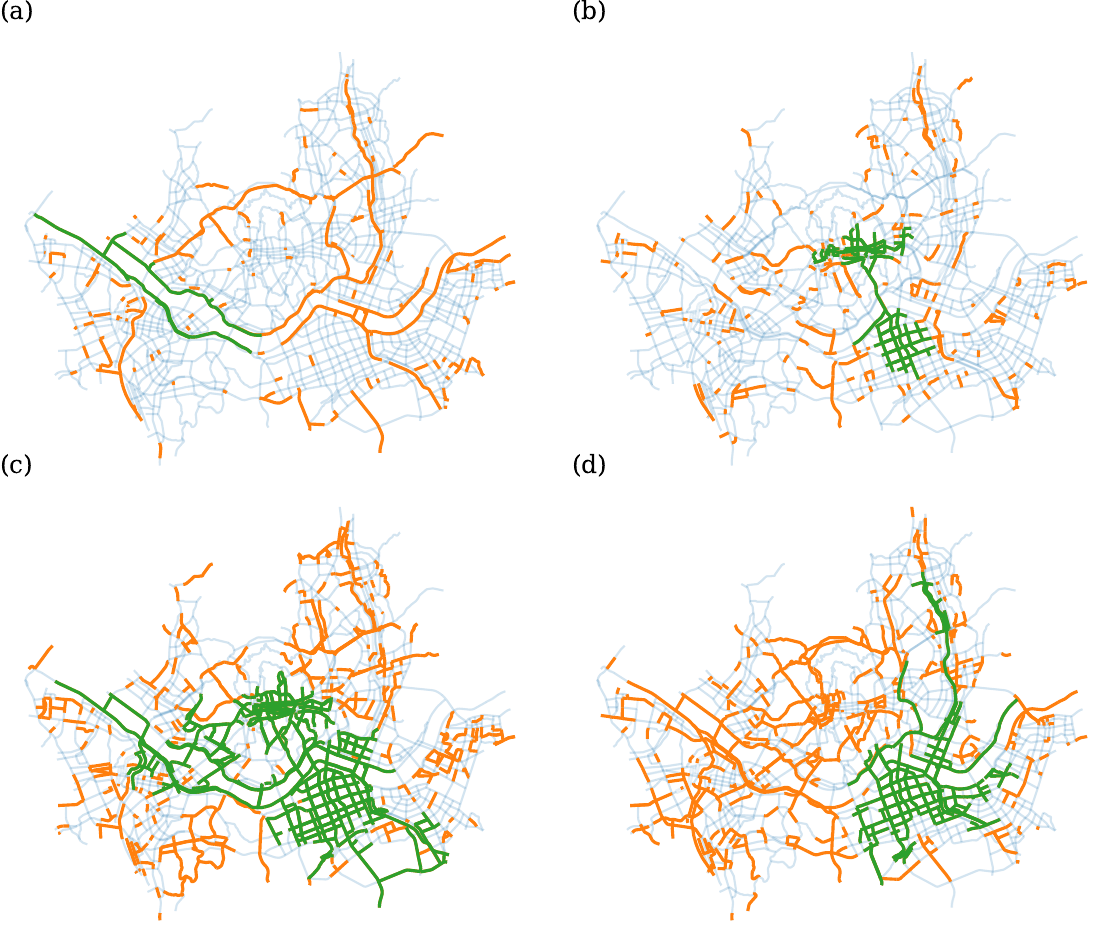}
    \caption{Spatial representation of congested flows and their largest weakly connected component.
    The date of the example data is Dec. 1st. 2020.
    We plotted the original Seoul road network as a guideline (thin blue), congested flows (orange), and the largest connected components (green).
    Each plot represents a representative time of day: (a) morning rush hour (07:00 A.M.), (b) lunchtime (01:00 P.M.), (c) before the evening rush hour peak (03:00 P.M.), and (d) after the evening rush hour peak (08:00 P.M.).
    }
    \label{fig:cong_snapshot}
\end{figure*}

\section{Spreading patterns of urban traffic congestion}

% importance of the structure of network in network dynamics
To study congestion propagation in urban traffic networks, we identified the congested traffic flows in the Seoul traffic network. 
We examined all 17,280 data points to get a set of congested flows in each snapshot of the Seoul traffic network. Fig.~\ref{fig:cong_snapshot} shows examples of spatial representations of congested flows and their largest weakly connected component for some representative times of Dec. 1st. 2020.

% Figure : ratio of congestion road and loops 
\begin{figure}
    \centering
    \includegraphics[width=\columnwidth]{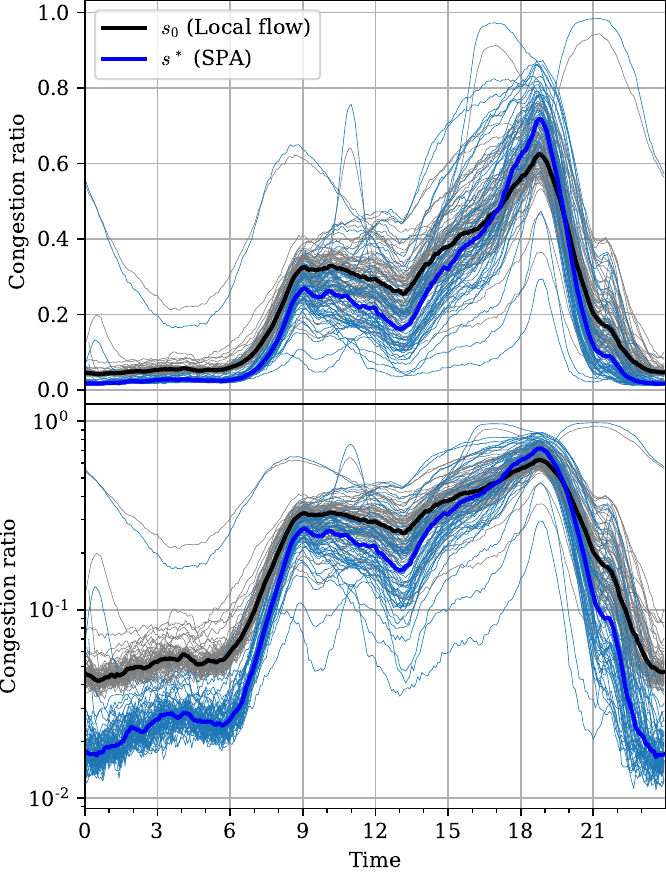}
    \caption{Daily evolution patterns of congested flows in the Seoul traffic network. 
    Each thin line represents a pattern of daily congestion ratio for different workdays,
    and the thick line shows the median as a guideline. %average.
    Blue and black colors represent the flow state determined by the Eq.~(\ref{eqn:initial}) (i.e., by considering only the flow itself) and congestion determined by the Eq.~(\ref{eqn:spa}) (i.e., by considering the neighbor flows together), respectively.
    The lower plot has the same data as the upper one but log scale on the y-axis.
    }
    \label{fig:c_ratio}
\end{figure}

% exponential behavior of rush hour in urban traffic congestion & tree-like structure
To understand the quantitative patterns of congestion spreading, 
we first check the evolution pattern of the number of congested flows $C(t)( = \sum_i c_i(t))$ in the Seoul traffic network (Fig.~\ref{fig:c_ratio}).
Each thin line represents the daily pattern of congestion evolution over 60 workdays, 
which shows a significant congestion growth during the morning rush hour and more severe congestion during the evening rush hour. 
We find a crucial structural pattern of evolving congestion, which is the exponential increase in the number of congested flows during the morning rush hour from 6 A.M. to 9 A.M. 
This exponential increase suggests that the spread of congestion during the morning rush hour has a tree-like structure as reported in other works~\cite{Saberi2020, chen2022quasicontagion}.
The spatial representation of congested flows during the morning rush hour also shows a tree-like structure (cf. Fig.~\ref{fig:cong_snapshot}(a)).
Due to construction costs, urban highways are often based on a tree structure, occasionally with a city-level ring structure~\cite{taillanter2023Evolution}. 
But, as observed in other works~\cite{zeng2019switch,kwon2023}, urban highways are vulnerable to congestion during rush hour.
Therefore, the observed tree structure of congested flows during the morning rush hour is likely to stem from the congestion of flows on urban highways.
 
% figure : Volume and surface of the congested LCC
\begin{figure}
    \centering
    \includegraphics[width=1.0\columnwidth]{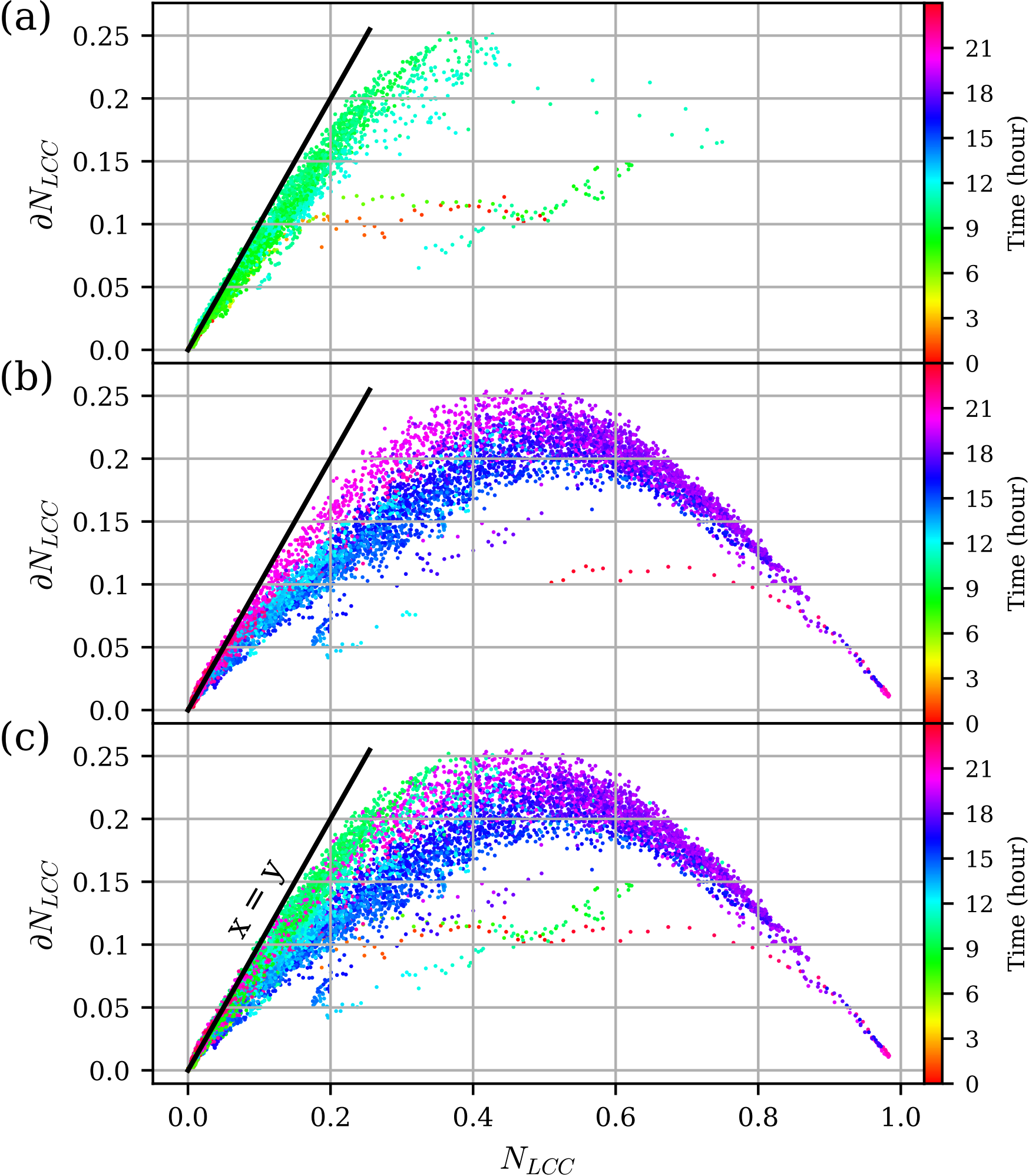}
    \caption{
    The relation between the sizes of the LCC and its boundary. 
    Each point represents each time snapshot (5 min interval), color-coded by the hour.
    One can see the difference between the two mainstreams of the daily pattern of the LCC, an increase and a decrease in congestion during the evening hours, which are colored blue and purple, respectively. 
    For the comparison between the morning and evening rush hour, we plotted the data based on several time zones: (a) 00:00 $\sim$ 12:00, (b) 12:00 $\sim$ 24:00, and (c) 00:00 $\sim$ 24:00.
    }
    \label{fig:surface}
\end{figure}

% relation between the largest connected component of urban traffic congestion and its neighboring flows
To address congestion spreading patterns in terms of connected clusters, 
we analyzed the weakly connected component (WCC) consisting of only congested flows in the Seoul traffic network. 
In particular, we trace not only the largest connected component (LCC) of congested flows but also its outer boundary which is a set of free-flowing flows connected to the LCC in the flow-to-flow network.
We defined the boundary $\partial N$ of a given cluster $N$ which is a set of flows as below,
\begin{equation}\label{eqn:surface}
    \partial N = \{i | i \notin N, j \in N, A_{ij} = 1\},
\end{equation}
where  $A_{ij}$ denotes the adjacency matrix of the Seoul traffic network.
This definition traces the candidates of flows that can be influenced by a given cluster $N$, meaning the total number of incoming neighbors of the cluster $N$.

% size proportionality and contagion process
Fig.~\ref{fig:surface} shows the relation between the sizes (i.e., number of flows) of the congested LCC and its boundary for each data point. One can see that the number of free-flowing flows that are connected to the LCC (i.e., the size of the boundary of the LCC) is proportional to the size of the LCC during the morning rush hour (green color in Fig.~\ref{fig:surface}).
This result is not only evidence for the tree-like structure of urban traffic congestion during the morning rush hour, but also an explanation of the exponential behavior of a growing pattern of urban congestion.

% separation and hysteresis
One can notice the separation of growth and relaxation patterns of the evening congestion, which are colored blue and purple in Fig~\ref{fig:surface}, respectively. These decoupled patterns of congestion LCCs reveal a hysteresis-like pattern in the evolution of urban congestion from growth to relaxation of congestion during the evening rush hour (cf. from Fig.~\ref{fig:cong_snapshot}(c) to (d)).
We can see that the growth pattern of the evening rush hour is sublinear (blue in Fig.~\ref{fig:surface}), whereas that of the morning rush hour and the relaxation pattern of the evening rush hour are near-linear (green and purple in Fig.~\ref{fig:surface}).
Because of the difference in the number of neighbors of the LCC, these hysteresis patterns suggest the structural shift of urban traffic congestion between the morning rush hour and the evening rush hour, and also the stability of such a structure in terms of the positive-feedback effect~\cite{angeli2004,liang2018feedback}.
Patterns of the morning rush hour (green in Fig.~\ref{fig:surface}) and the relaxation process of the evening rush hour (purple in Fig.~\ref{fig:surface}) are not much different in terms of the number of neighboring free-flowing flows of the congestion LCC (cf. Figs.~\ref{fig:cong_snapshot}(a) and (d)).
While such patterns of congestion can be explained by tree-like structures, the emerging clusters of congested flows during the evening rush hour have fewer neighbors on their boundaries than during its relaxation process or the morning rush hour.
To summarize, these results indicate that there is a topological shift between the morning and evening rush hours in the largest cluster of congested flows and hysteresis in the evolutionary pattern of urban congestion.
However, these results are not sufficient to explain why congestion is much larger in the evening or to explain where the structural differences between the morning and evening rush hours come from.

\begin{figure}
    \centering
    \includegraphics[width=\columnwidth]{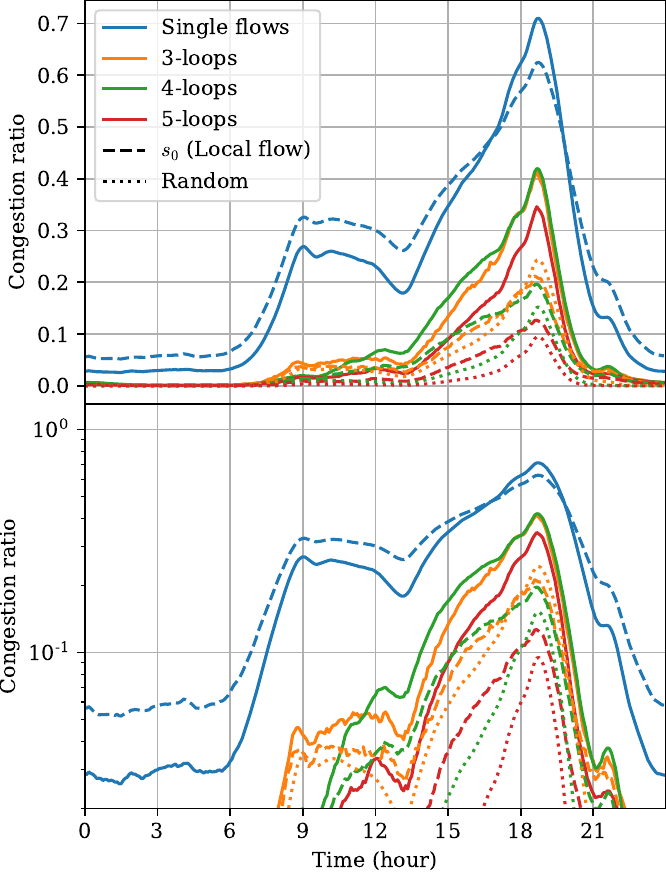}
    \caption{Daily evolution patterns of congested flows and loops in the Seoul traffic network. 
    Each solid line represents the workday average of the temporal evolution of the congestion ratio of each flow and loop of 3, 4, and 5 flows combined. 
    The results of congestion identification based on a local flow ($s_0$) are represented as the dashed line.
    As a guideline, each dotted line shows the probability that each k-flows loop gets congested with a given congestion ratio of single flows, which is calculated as $p^k$ where $p$ and $k$ denotes the overall congestion probability of $s_0$ and the number of traffic flows in loops, respectively.
    The lower plot has the same data as the upper one but log scale on the y-axis.
    }
    \label{fig:loops}
\end{figure}

% loop structure and positive feedback
To address the above questions, we focused on a special topological feature, \textit{loops}, consisting of congested flows.
A loop (or cycle) of congested flows is a set of simultaneously congested flows that form a closed path in the network.  
These loops, especially small loops, are very important structures in the network dynamics that determine how one's influence is reflected back to oneself.
For example, the most important characteristic that determines a tree structure is the absence of the loop structure in the network.
Moreover, when we look at congestion as a failure, these loops can be seen as a cycle of failures, which represents a kind of feedback effect where the effects of one's own failure cascade back to oneself. 
Therefore, we investigated how urban congestion is structurally different between morning and evening through these loops. 

% loop detection and loop congestion
We have found the set of $k$-loops $L_k$ which are made up with $k$ flows in the traffic network,
and calculated the congestion indicator $c_l(t)$ of a loop $l_i$ at time $t$ as below,
\begin{equation}
    c_l(t) = \prod_{j\in \{l_i\}}^k \Theta(-s^*_j(t)), \label{eq:6}
\end{equation}
where $\Theta(\cdot)$ denotes the Heaviside step function. 
This congestion indicator for loops is 1 only if all the consisting flows are in a congested state (otherwise 0).
We calculated the above indicator only for the loops with 3, 4, and 5 flows because the larger loops are less important in terms of the feedback effect.
We investigated all 3-, 4-, and 5-flow loops in the Seoul traffic network in a brute-force manner. In Appendix B, Fig. 8 shows a schematic of a 4-loop and Fig. 9 shows a real-space representation of all possible 3- and 4-loops in the Seoul road network.

% result & meaning
Fig.~\ref{fig:loops} shows the time evolution of the number of $k$-loops $C_k(t)(=\sum_{l\in L_k} c_l(t))$ over a day.
As we expected above, only less than $5\%$ of loops are congested in the morning periods from 6 A.M. to 9 A.M., while the congestion ratio of individual flows is over $20 \%$.
This absence of small loops is clear evidence of the tree structure in the evolution pattern of urban congestion during the morning rush hour.
However, after roughly 1 P.M., small-size loops of congested flows emerge drastically so that the congestion ratio for each loop reaches nearly 40\% or more about 7 P.M., which is not shown by the congestion identification based on a single flow (dashed lines in Fig.~\ref{fig:loops}).
These results suggest that despite the fact that the overall traffic volumes during the morning and evening commutes are not significantly different~\cite{TOPIS}, there are significant differences in traffic flow dynamics between the two time periods, and that the difference in the congested loops is responsible for the differences in traffic flow dynamics. 
On the other hand, after the number of congested flows reaches a peak, these loops decrease sharply, so that the relaxation pattern in the evening commute shows a tree structure similar to the spreading pattern in the morning commute, as expected above.

\begin{figure}
    \centering
    \includegraphics[width=\columnwidth]{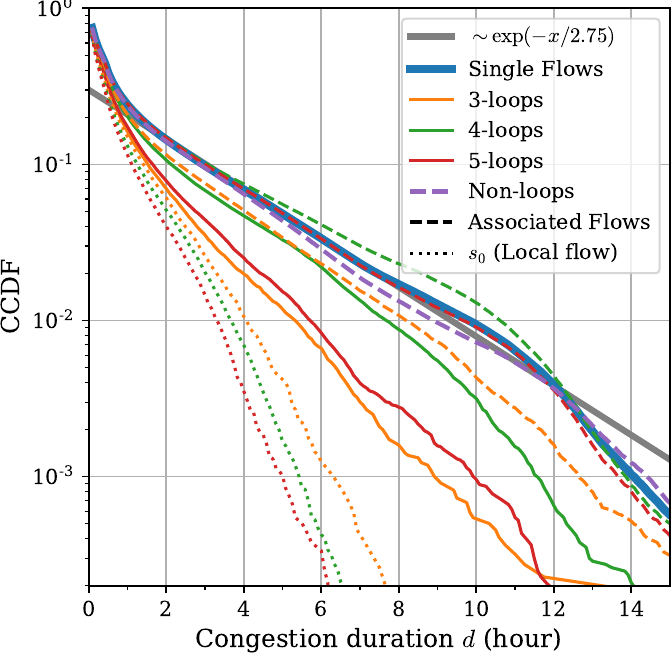}
    \caption{The distribution of congestion duration of loops and individual flows. 
    Each line represents the decay pattern of each type of congested loop. 
    The blue solid line indicates the congestion that was appeared in every single traffic flow and its tendency shown with the grey solid line as a reference.
    The yellow, green, and red lines indicate the congestion duration of loops with 3, 4, and 5 flows, respectively.
    To investigate the impact of congested loops on a single flow, we categorized flows by the loop they belong to.
    The dashed line describes the congestion duration on traffic flows that are classified as associated loops, and the purple line shows exceptional flows that do not consist of loops.
    % null model 설명 -> 본문으로 
    For comparison, we generate random congestion in each loop by shuffling only the spatial organization of the flows, leaving the structure of the network unchanged.
    Each dotted line represents the congestion duration distribution for each loop with the random flow configuration.
    }
    \label{fig:duration}
\end{figure}

% stability analysis of small loops of urban traffic network
To understand the impact of the loop structure on urban congestion propagation, the distribution of congestion duration $d$ was obtained by calculating the consecutive time of congested states for each traffic object (i.e., flows and loops) $i$, which is described as follows. 
First, let us consider the congestion starting point indicator $o_i(t)$ as below,
\begin{equation}
    o_i(t)= c_i(t)\cdot(1-c_i(t-1)),
\end{equation}
where $c_i(t)$ denotes the congestion indicator of a traffic object $i$ defined in Eq.~(\ref{eq:6}).
This indicator $o_i(t)$ shows 1 if congestion emerges at time $t$, else 0.
After that, we considered the congestion length $l_i(t)$ to be the farthest time shift $\tau$ that represents the continuous congestion of traffic object $i$ from time $t$,
\begin{equation}\label{eqn:cong_length}
    l_i(t) = \sum^{T-t}_{\tau =0} \prod^{\tau}_{\Delta t=0} c_i(t+\Delta t),
\end{equation}
where $T$ denotes the total number of data points.
If $c_i(t+\tau)=0$ once, then for any $\tau$ greater than that, the term inside of a sum will always be zero.
So, this calculation allows us to know the maximum length of consecutive congestion from time $t$.
With above indicators, we calculated the congestion duration $d_i(t)$, which can be written as below, 
\begin{align}\label{eqn:cong_duration}
    d_i(t) = 
    \begin{cases}
        0, & (\text{if}\quad  o_i(t) = 0)\\
        l_i(t),    & (\text{otherwise,})\\
    \end{cases}
\end{align}
Finally, we collected all the positive congestion duration that occurred in a certain set $S$ of traffic objects, which can be formulated as follows,
\begin{equation}
    D(S) = \{d_i(t)| d_i(t) >0, i \in S\}.
\end{equation}
We prepared a set of traffic objects based on various categories (e.g., loops by the number of flows, flows by associated loops) to investigate the impact of loop structure on congestion.
In addition, to remove spatial correlation, we shuffled only the flow configuration, leaving the structure of the road network intact, and examined the congestion on the loops of that network.
In this way, we can obtain a congestion duration distribution and its characteristic decay time, which represents the persistence of urban congestion for each traffic object.

Fig.~\ref{fig:duration} shows the complementary cumulative density function of congestion duration distribution of each traffic object. 
In general, all loop structures which are represented as solid lines were found to be more persistent than the shuffled ones which are represented as dotted lines.
Especially, even though the congestion of a loop is less probable than that of a single flow, for 4-flow loops (green line), the characteristic time is similar to a single flow.
This result suggests that the positive feedback of the loop structure makes the structure more persistent. 
Once such a cycle of congestion in traffic networks is created, 
it would not disappear easily and make things worse by disrupting neighboring traffic flows.
This is even more pronounced when we separate the distributions based on which loop a flow belongs to, 
and find that flows belonging to a loop of four flows (green dashed line) experience significantly longer congestion than those that do not.
In addition, traffic flows that do not form small loops (purple dashed line) show a shorter characteristic time than flows in 4- or 5-flow loops (green and red dashed lines, respectively), meaning that congested flows in a tree structure experience shorter congestion than congested flows in a loop structure.

% meaning and summary of sec.3. : superseded effect of loop structure -> Discussion? or not?
All these results indicate that the essential patterns of urban congestion propagation are the tree and loop structures.
The tree structure that emerges during the morning rush hour and the relaxation part of the evening rush hour tells us the basic spreading pattern of urban congestion, which is the diffusion-like (or contagion-like) process.
We expect that this pattern stemmed from the structure of the highway network, which serves as a long-range connection through the city and is designed for efficiency. % Barthelemy 2023
The other can be explained by the stacking process of urban congestion, the small loop structure in which congestion resulting from the morning rush hour is not relaxed during the midday, thus exacerbating urban congestion during the evening rush hour.
Flows that make up a small loop composed of 4 or 5 flows are identified as having a longer congestion duration than flows that are not part of the loop.

\section{Discussion}
% Summary
In summary, we developed a systematic framework to analyze congestion spreading in empirical data by viewing traffic flows in cities as network flows, defining congestion as an anomalously low functional state of these flows. 
Our framework enables us to determine the functional state of traffic flows in urban traffic networks by collapsing the behavior of various types of traffic flows onto a single  type of  statistical distribution and taking into account the functional states of their neighboring flows.
As a result, we found the tree structure in congestion evolution patterns during the morning rush hour observed in the exponential growth in the number of congested flows, the near-linear relation between the size of the largest connected component of congested flows and the size of its boundary, and the lack of small loops during the morning rush hour.
On the other hand, we observed a significant increase in the number of small-size loops of congested flows during the evening rush hour. We observed that these loops are quite persistent as they represent the feedback effect of urban congestion.

% Meaning
Our findings suggest that evaluating the dynamic state of nodes in networks by taking into account the state of their neighboring nodes is helpful in providing a clearer picture of dynamical processes on networks. In the case of traffic dynamics, by propagating the information of each flow's functional state, we are able to reconstruct the structural patterns of congestion spreading, such as trees and loops.

% limitation
Although our framework provides an effective tool for understanding urban congestion, it also has some limitations.
We only identify congestion as the failure of each traffic flow based on an estimation of an undisturbed velocity distribution. 
So, if the velocity distribution of a traffic flow is already too slow that congestion cannot be distinguished by our estimation, we cannot identify the impact of this congestion in the data.
Moreover, our findings are still limited to the phenomenon in one city, Seoul, and need to be further validated with data from various cities.
Lastly our results may depend on the values of parameters $h$ and $J$ in Eq. (4) although no qualitative differences were observed when we checked the results with other parameter values (See Appendix C for the details).  
Despite these limitations, it can be seen that the algorithm is powerful in revealing various aspects of traffic dynamics in urban traffic networks.

% Application, Future Work & Outro
As an immediate follow-up study, we will investigate urban congestion in other cities to identify universal patterns in the spreading process of urban traffic congestion. 
We will also apply this algorithm to solve other collective phenomena that occur in networked systems and understand the spatio-temporal patterns of these phenomena to reveal the relationship between structure and dynamics. 
We hope that understanding the circular effects of urban traffic congestion will help traffic engineers and road network designers solve the socioeconomic problems of urban congestion by alleviating severe traffic congestion in large cities.

\begin{acknowledgments}
This work was supported by the 2021 Research Fund of the University of Seoul. The authors thank the Seoul Metropolitan Government for sharing the Seoul road traffic data. The authors also acknowledge the Urban Big data and AI Institute of the University of Seoul supercomputing resources (http://ubai.uos.ac.kr) made available for conducting the research reported in this paper.
\end{acknowledgments}

\appendix 

% LSA
\section{Stability analysis of state propagation}
To estimate the stability of the converged state $s^*_i$, we examine the small perturbation $\delta_i$ at each node $i$ and calculate its evolution.
    % Perturbation analysis of Eq (4) 
    Let $s^*_i$ be a converged state of a given dataset $z_i$. We can calculate the propagation of the small perturbation $\delta_i$ on $s^*_i$.
Then, a propagated fluctuation $\delta'_i$ can be written as below,
\begin{equation}\label{eqn:pertubation}
\begin{split}
    s^*_i + \delta'_i  &= \tanh \left( \sum_j J \frac{A_{ij}(s^*_j +\delta_j)} {k_i} +z_i +h \right) \\
\end{split}
\end{equation}
where $k_i(=\sum_j A_{ij})$ is out-degree of flow $i$. Because of the small perturbation condition, the right-hand side can be expanded by $\delta_j$ at $s^*_j$ by using the relation, $\tanh'(x) = 1- \tanh^2 x$,
\begin{equation}\label{eqn:expansion}
\begin{split}
    \delta'_i &\simeq (1 - (s^*_i)^2)J\overline{\delta_i} + \left(-s^*_i (1 - (s^*_i)^2)\right) (J\overline{\delta_i})^2+\cdots,\\ \overline{\delta_i} &= \sum_j \frac{A_{ij}\delta_j} {k_i} ,
\end{split}
\end{equation}
where $\overline{\delta_i}$ means the average fluctuation of outgoing neighbors of flow $i$. Higher-order terms of $J\overline{\delta_i}$ contain higher-order of $s^*_i$ or $(1-(s^*_i)^2)$ which are always smaller than 1.
So, the size of the propagated fluctuation $\delta'$ always smaller than one of the original perturbation $\delta$ when $(1 - (s^*_i)^2)J <1$ holds ($J\leq 1$). 
Therefore, the iteration of Eq.~\ref{eqn:spa} always converged to $s^*$ in that condition.

In addition, the initial state $s^0_i$ is the same as the condition $\delta_i = - s^*_i$ in Eq.~\ref{eqn:pertubation} which can be written as below,
\begin{equation}
    s^0_i = s^*_i + \delta^0_i = \left( \sum_j J \frac{A_{ij}(s^*_j +(-s^*_j))} {k_i} +z_i +h \right)
\end{equation}
So, we can calculate the difference $\delta^1_i$ between the first iteration $s^1_i$ and the converged state $s^*_i$ as below,
\begin{equation}
\begin{split}
    \delta^0_i &\simeq (1 - {s^*_i}^2)(-J\overline{s^*_i}) - \left(s^*_i (1 - (s^*_i)^2)\right) (-J\overline{s^*_i})^2+\cdots,\\
    \overline{s^*_i} &= \sum_j \frac{A_{ij}s^*_j} {k_i} ,\quad
\end{split}
\end{equation}
Because $s^*_i \in [-1,1]$, our initialization method (i.e. Eq.~\ref{eqn:initial}) is stable to trace the converged state $s^*_i$ under the condition $J\leq 1$.

\section{Real-space representation of congested loops}

\begin{figure}[H]
    \centering
    \includegraphics[width=0.5\linewidth]{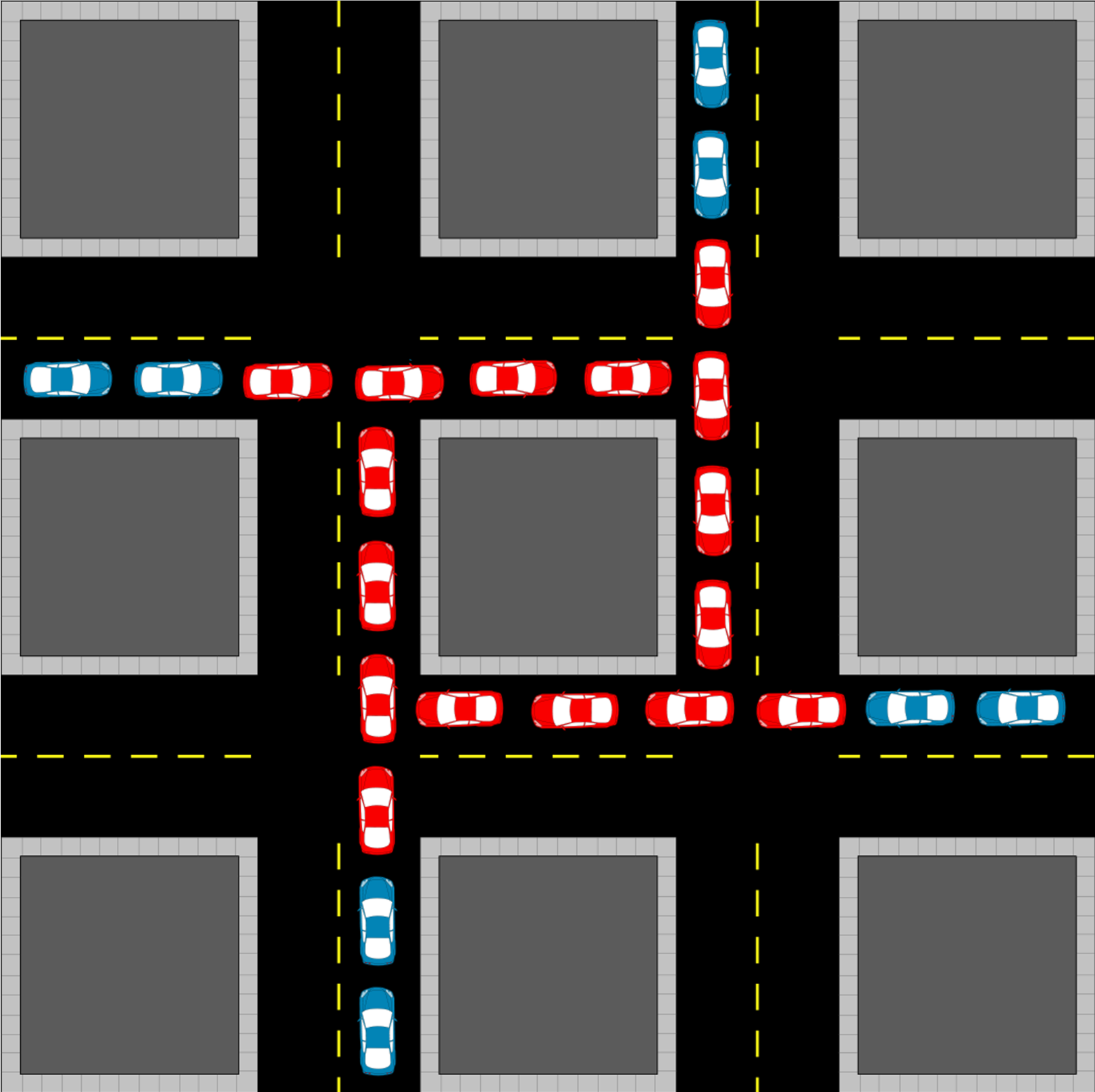}
    \caption{Schematic of a 4-loop composed of 4 congested flows (red).}
    \label{fig:schematics}
\end{figure}

\begin{figure}[t]
    \centering
    \includegraphics[width=\linewidth]{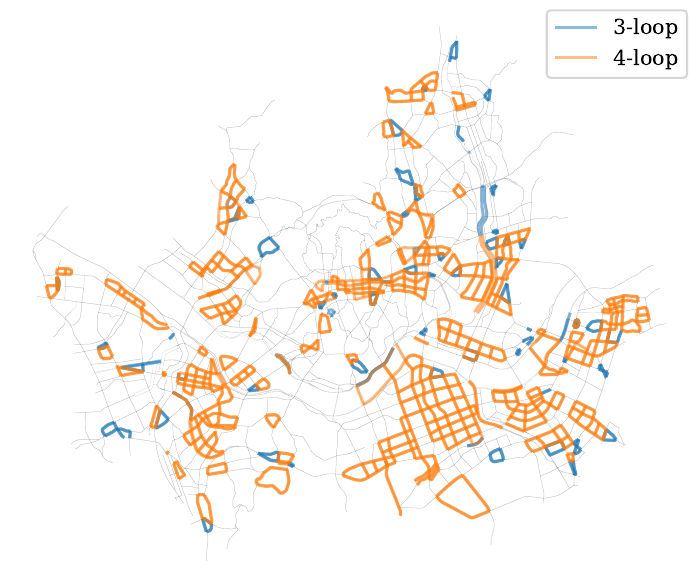}
    \caption{Real-space representation of all possible 3- and 4-loops of congested flows in Seoul road network. The black thin line represents the underlying Seoul Road network. Blue and Orange show all 3- and 4-loops of congested flows in the Seoul road network, respectively. Note that the bidirectional road might be overlapped in the real-space representation. }
    \label{fig:loopRS}
\end{figure}

\section{Impacts of different parameter values}
\begin{figure*}
    \centering
    \includegraphics[width=\textwidth]{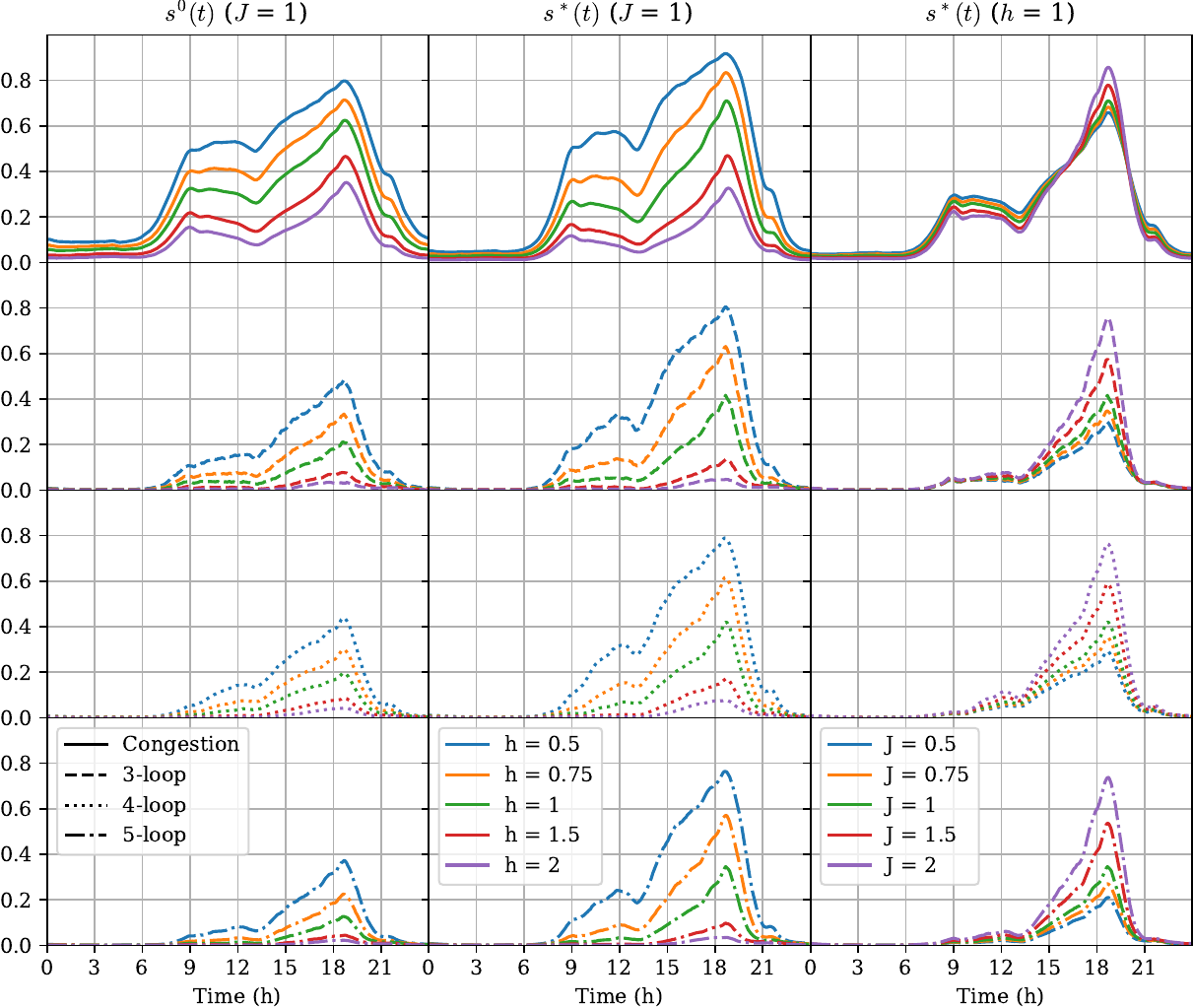}
    \caption{Daily evolution patterns of congested flows and loops for different combinations of parameters $J$ and $h$. The left and middle columns show the result of choosing different values for the congestion threshold $h$ with different colors, while the calibration strength $J$ is fixed to 1. The right column shows the result of choosing different values for the calibration strength $J$ with different colors, while the congestion threshold $h$ is fixed to 1. Because the initial state $s^0(t)$ is determined only by the congestion threshold $h$, the green color in the left column corresponds to the initial state of the right columns. }
    \label{fig:varioush}
\end{figure*}

First, we tested various values of the congestion threshold $h$, given the calibration strength $J$ = 1 in Fig.~\ref{fig:varioush}.
The tendencies of different congestion thresholds are similar to one another, while the absolute fraction decreases according to the increasing threshold as shown in Fig.~\ref{fig:varioush}.

Let us start from the original normalized data $s^0_i(=s^*_i (J=0))$ and consider the sorted dataset $z_i(t)$ by the increasing order.
In this case, the congestion threshold $h$ determines only the number of congested flows($s^0_i(t) = \tanh(z_i(t) + h)$), leaving the order of congestion still unchanged.

Second, we also tested various values of the calibration strength $J$, given congestion threshold $h$ = 1 in Fig.~\ref{fig:varioush}.
The tendencies of different calibration strengths are similar to one another, while the absolute fraction decreases according to the decreasing strength as shown in the third column in Fig.~\ref{fig:varioush}.
The calibration strength $J$ determines how strong the influence of local traffic states is, which emphasizes the structure composed of flows whose states are strong. 
In other words, the converged state $s^*_i$ of the flow $i$ whose state is stronger than $|z_i + h| > |J|$ is not affected by the calibration. 
Therefore, choosing $J$ determines the boundary of a strong state and the ambiguous state which may contain some fluctuations (e.g. errors).

\bibliography{CSNS2306}

\end{document}